\documentclass{article}

\usepackage{microtype}
\usepackage{graphicx}
\usepackage{subfigure}
\usepackage{booktabs} 

\usepackage{hyperref}

\usepackage[accepted]{icml2024}

\usepackage{amsmath}
\usepackage{amssymb}
\usepackage{mathtools}
\usepackage{amsthm}

\usepackage[capitalize,noabbrev]{cleveref}

\theoremstyle{plain}

\theoremstyle{definition}

\theoremstyle{remark}

\usepackage[textsize=tiny]{todonotes}

\icmltitlerunning{A Likelihood-Based Generative Approach for Precipitation Downscaling}

\begin{document}

\twocolumn[
\icmltitle{A Likelihood-Based Generative Approach for Spatially Consistent Precipitation Downscaling}



\icmlsetsymbol{equal}{*}

\begin{icmlauthorlist}
\icmlauthor{Jose González-Abad}{1}
\end{icmlauthorlist}

\icmlaffiliation{1}{Instituto de Física de Cantabria (IFCA), CSIC-Universidad de Cantabria, Santander, Spain}

\icmlcorrespondingauthor{Jose González-Abad}{gonzabad@ifca.unican.es}

\icmlkeywords{Machine Learning, ICML}

\vskip 0.3in
]



\printAffiliationsAndNotice{}  

\begin{abstract}
Deep learning has emerged as a promising tool for precipitation downscaling. However, current models rely on likelihood-based loss functions to properly model the precipitation distribution, leading to spatially inconsistent projections when sampling. This work explores a novel approach by fusing the strengths of likelihood-based and adversarial losses used in generative models. As a result, we propose a likelihood-based generative approach for precipitation downscaling, leveraging the benefits of both methods.
\end{abstract}

\section{Introduction}

Global Climate Models (GCMs) simulate the spatio-temporal evolution of climate by numerically solving the set of physical equations describing its constituent components and interconnections \cite{chen_framing_2021-1}. By running these models under various emission scenarios, it is possible to generate future projections under climate change conditions \cite{o_scenario_2016}. However, GCMs have coarse resolutions due to computational and physical constraints, limiting their use in regional-scale studies.

Statistical Downscaling (SD) attempts to overcome this limitation by modeling the relationship between the coarse (low-resolution) and regional (high-resolution) scales. Under the Perfect Prognosis (PP) approach \cite{maraun_statistical_2018}, an empirical link is established between a set of large-scale synoptic variables (representing the state of the atmosphere) and the local variables of interest (e.g., precipitation). This link is learned via a statistical model using observational datasets. The model is then applied to the large-scale variables from a GCM to obtain the corresponding regional projections for future scenarios.

Recently, Deep Learning (DL) has emerged as a promising technique for Perfect Prognosis Statistical Downscaling (PP-SD), given its ability to model non-linear relationships and handle spatial data \cite{goodfellow_deep_2016}. The resulting DL-based regional projections have proven useful for several climate change applications, even forming the first continental-wide contribution of a PP-based technique to the CORDEX initiative \cite{bano_downscaling_2022}.

Unfortunately, regression-based DL models tend to focus on capturing the expected value of the output distribution, thus leading to the underrepresentation of extremes. This can be problematic for variables such as precipitation, highly characterized by these events (e.g., heavy rainfalls). To address this, recent DL models for PP-SD maximize the likelihood with respect to a explicit distribution reflecting the dynamics of precipitation \cite{bano_downscaling_2022}. However, because these distributions are learned independently for each grid-point at the regional scale, the resulting projections may lack spatial consistency when sampling \cite{gonzalez_use_2021}.

In this work, we explore generative models to address this issue. For the first time, we combine a likelihood-based training with a conditional Generative Adversarial Network (cGAN) for the downscaling of precipitation. Through our experimental setup, we show how the proposed approach enables the DL model to sample spatially consistent precipitation fields, while allowing defining a explicit probability distribution over the target data.

\section{Background}

\subsection{Deep Learning for Precipitation Downscaling}

Inspired by advances in the Super Resolution (SR) field \cite{wang_deep_2020}, recent studies have explored SR models for the downscaling of precipitation \cite{vandal_deepsd_2017,cheng_reslap_2020,passarella_reconstructing_2022,sharma_resdeepd_2022}. In this context, SR models aim to establish a link between the coarse and regional versions of a specific field. However, using a surface variable from a GCM often results in reproducing the regional biases caused by its coarse resolution, limiting their effectiveness for climate downscaling.

For this reason, in the climate context PP-SD models constitute the standard, as they rely on large-scale synoptic variables representing the state of the atmosphere, which are properly reproduced by the coarse resolution of GCMs. Several architectures have been explored such as recurrent networks \cite{misra_statistical_2018}, a combination of convolutional and dense layers \cite{pan_improving_2019,bano_configuration_2020} and fully-convolutional \cite{adewoyin_tru_2021,quesada_repeatable_2022}, some of them even generating projections in future scenarios \cite{bano_suitability_2021,soares_high_2023}. Among these architectures, the U-Net \cite{ronneberger_unet_2015} has recently shown promising results \cite{quesada_repeatable_2022, adewoyin_tru_2021}, even in related fields such as emulation \cite{doury_regional_2023,doury_suitability_2024}.

\subsection{Extreme Precipitation}

Due to the dynamics of precipitation, which typically adheres to exponential probability distributions, and its non-continuous nature (occurrence and amount), regression-based DL models encounter difficulties in accurately modeling it. This often leads to significant issues, such as underestimation of extreme precipitation events \cite{rampal_enhancing_2024}.

To address this challenge, and drawing inspiration from previous work \cite{dunn_occurrence_2004,cannon_probabilistic_2008}, authors in \citep{bano_downscaling_2022} train a DL model by minimizing the Negative Log-Likelihood (NLL) of a Bernoulli and gamma distributions for the occurrence and amount, respectively. By working under this assumption, which aligns with the dynamics of precipitation \cite{williams_modelling_1997}, they are able to model the whole distribution, including the extremes. Its success has led to its application across various regions \cite{sun_statistical_2021,rampal_high_2022,kheir_improved_2023,hosseini_improving_2024}, making it the most extended DL-based PP-SD model. Unfortunately, this approach models a different probability distribution for each of the grid-points forming the downscaled variable, resulting in spatial inconsistency when sampling from these distributions, leading to unrealistic projections \cite{gonzalez_use_2021}.

\subsection{Generative Precipitation Downscaling}

Recently, Generative Adversarial Networks (GANs) \cite{goodfellow_generative_2014} have attracted significant attention in the SD field. Unlike regression-based DL models, which generally minimize loss functions aiming at capturing the mean (e.g., Mean Squared Error, MSE), GANs minimize an adversarial loss which encourages the generator to better reproduce the underlying distribution of the data in order to fool the discriminator. This leads to improved reproduction of extremes and finer details in precipitation downscaling \cite{rampal_enhancing_2024}. In addition, GANs allow computing ensembles of predictions, although unlike the NLL approach, these are computed by passing noise as input to the generator, not by explicitly modeling the corresponding distribution. In the context of PP-SD, GANs have not been explored yet, although they show promise in related areas such as SR downscaling \cite{leinonen_stochastic_2020,cheng_generating_2020}, meteorological downscaling \cite{price_increasing_2022,harris_generative_2022} and emulation \cite{rampal_robust_2024}.

\section{Experimental Framework}

\subsection{Region of Study and Data}

In this work, we focus on daily precipitation downscaling over a domain centered on the Alps (37.6°N-50.4°N and 3.6°E-16.4°E), a region of interest due to its prominent orography, which significantly influences local precipitation. As we frame this study within the PP approach, we rely on the ERA5 reanalysis dataset \cite{hersbach_era5_2020} (quasi-observational) at $1^\circ$ resolution for the large-scale variables (predictors) and on the observational dataset E-OBS \cite{cornes_ensemble_2018} for the local-scale variable (predictand) at $0.1^\circ$ resolution. Following previous works \cite{bano_downscaling_2022,soares_high_2023}, we select as predictors the air temperature, specific humidity, geopotential height, and the meridional and zonal wind components at 500, 700, and 850 hPa.

\subsection{Standard Deep Learning Models}
\label{sec:periods}

Following the aforementioned recent advances in PP-SD, we rely on the U-Net, a fully-convolutional model. Specifically, we adhere to the implementation details described in \cite{doury_regional_2023}. We train two different versions of this model: U-Net (MSE) and U-Net (NLL). The former minimizes the MSE, whereas the latter minimizes the NLL of a Bernoulli-gamma distribution, as proposed in \cite{bano_downscaling_2022}. While U-Net (MSE) directly computes the downscaled precipitation, U-Net (NLL) computes, for each grid-point in the predictand, the parameters $p$, $\alpha$ and $\beta$ defining the corresponding Bernoulli-gamma distribution. Consequently, for NLL-based models, the final prediction corresponds to a random sample from the modeled distributions.

We divide the observational dataset into a training (1980-2010) and a test (2011-2022) period. These models are trained using the Adam optimizer \cite{kingma_adam_2014} with a learning rate of $10^{-4}$ and a batch size of 64.

\subsection{Likelihood-Based Generative Approach}

The main contribution of this work involves cGAN models \cite{mirza_conditional_2014} for PP-SD. Unlike standard GANs, cGANs allow conditioning the generation process on specific data by feeding it to both the generator and discriminator. In addition to the adversarial loss, cGANs also minimize a content loss that ensures that conditionally generated samples align with the true target values. In the context of precipitation downscaling, combining these loss functions allows the generator to produce precipitation fields that both fool the discriminator and accurately reproduce the daily precipitation conditioned on the specific large-scale synoptic state \cite{harris_generative_2022,rampal_robust_2024}.

To represent the state-of-the-art in generative precipitation downscaling, we follow \citep{ravuri_skilful_2021,harris_generative_2022} and implement what we denote as cGAN (MSE), as this model relies on the MSE as its content loss. However, to avoid the so-called \textit{blurry effect}, where the model smooths out fine details and produces predictions resembling the mean, the content loss is applied to the mean of an ensemble of predictions. Note that, although the content loss is computed over an ensemble, at inference time the prediction of this model corresponds to a single sample.

The main contribution of this work is the introduction of the cGAN (NLL) model, which combines likelihood-based training with the cGAN framework. Unlike other approaches, this model uses the NLL of the Bernoulli-gamma distribution as the content loss. As a result, the generator produces a set of probability distributions, similarly to U-Net (NLL). However, by passing a random sample from these distributions to the discriminator, the generator is forced to improve the spatial consistency to minimize the adversarial loss, as spatially inconsistent precipitation fields are easily detectable by the discriminator. Therefore, adversarial training should lead the generator to learn spatially-aware distributions across grid-points in the downscaled field, addressing the drawbacks of standard likelihood-based DL models.

For both cGAN (MSE) and cGAN (NLL) models, we use the same architecture for the generator as that used for the U-Net (MSE) and U-Net (NLL) models. For the discriminator, we implement a fully convolutional network that processes both the large- and regional-scale data through a series of convolutional and dense layers. Following previous works \cite{leinonen_stochastic_2020, harris_generative_2022, rampal_robust_2024}, we rely on the Wasserstein formulation of GANs \cite{arjovsky_wasserstein_2017} with a gradient penalty term \cite{gulrajani_improved_2017}. This training framework is popular due to its theoretical properties, such as the possibility to train the discriminator to optimality. For both cGAN models, and following \cite{arjovsky_wasserstein_2017}, we use RMSprop as the optimizer with a learning rate of $10^{-5}$ and a batch size of 64. The generator and discriminator are trained in an adversarial manner, with the discriminator being updated five times for each update of the generator. The training and test sets cover the years detailed in Section \ref{sec:periods}.

\section{Results}

\begin{figure}[ht]
\vskip 0.2in
    \begin{center}
        \centerline{\includegraphics[width=0.95\columnwidth]{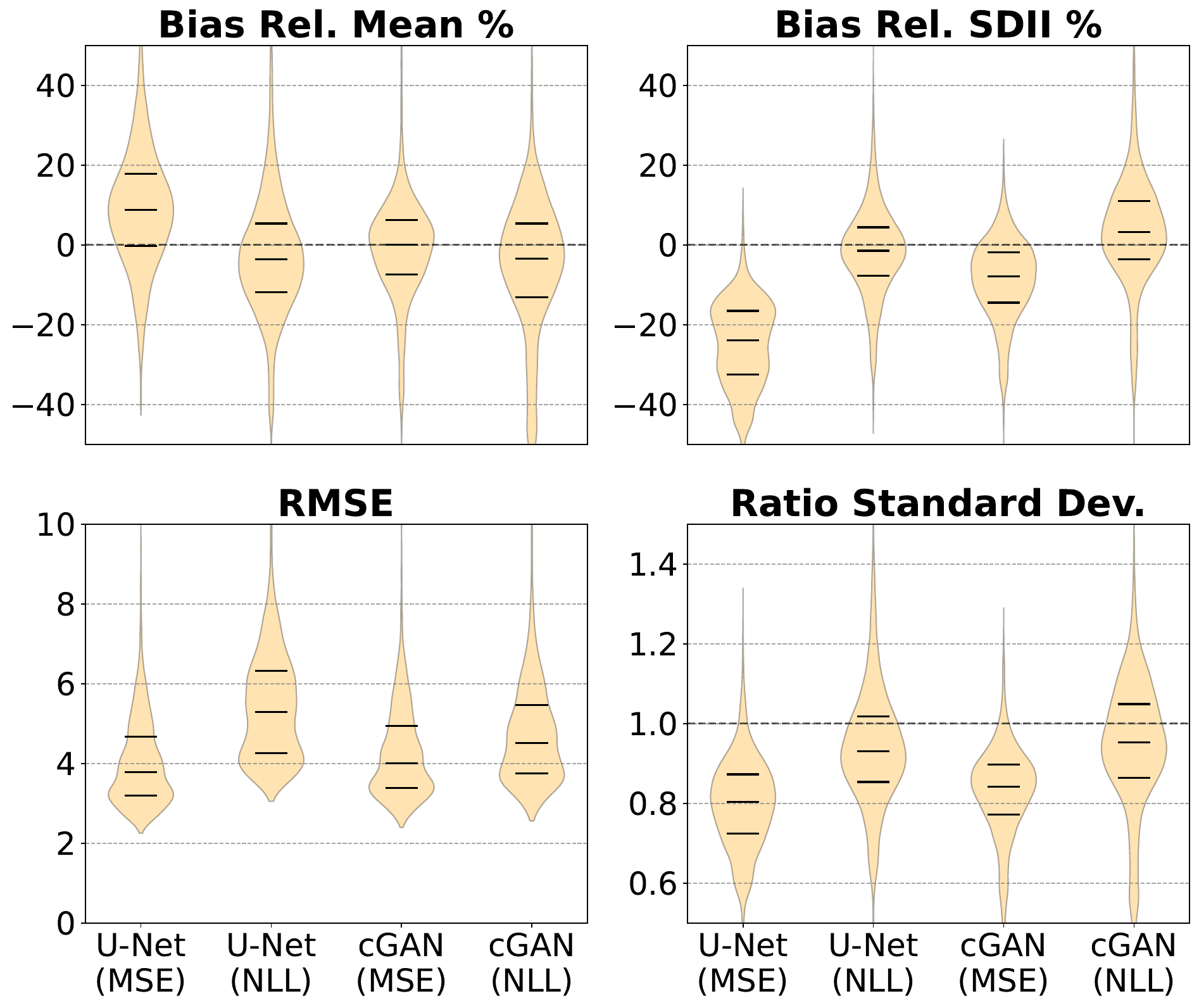}}
        \caption{Violin plot showing the results for four different metrics computed on the test set: the relative bias of the mean and the SDII, the RMSE, and the ratio of standard deviations. Each metric displays the results corresponding to the different DL models intercompared: U-Net (MSE), U-Net (NLL), cGAN (MSE), and cGAN (NLL).}
        \label{fig:violin}
    \end{center}
\vskip -0.2in
\end{figure}

Figure \ref{fig:violin} shows the violin plot for four different metrics computed on the test set: the relative bias of the mean, the relative bias of the Simple Daily Intensity Index (SDII), which corresponds to the precipitation amount for rainy days ($\geq 1$mm), the Root Mean Square Error (RMSE), and the ratio of standard deviations. These metrics are shown for the four intercompared models: U-Net (MSE), U-Net (NLL), cGAN (MSE), and cGAN (NLL). 

Besides the U-Net (MSE) which overestimates it, all models reproduce the mean. However, for the SDII, which represents values further from the mean, MSE-based models, particularly U-Net, tend to underestimate it, whereas NLL-based models are able to reproduce it accurately. As expected, MSE-based models reveal superior performance for the RMSE due to their relation with the minimized loss function. Although NLL-based models are expected to exhibit higher RMSE due to sampling-induced errors, cGAN (NLL) achieves comparable performance to MSE-based models. In terms of the ratio of standard deviations, NLL-based models show superior performance, primarily attributed to sampling from the Bernoulli-gamma distribution. However, the cGAN (MSE) model, despite sampling, falls short in accurately reproducing this aspect, which may indicate that explicitly defining the distribution of precipitation is key.

\begin{figure}[ht]
\vskip 0.2in
    \begin{center}
        \centerline{\includegraphics[width=\columnwidth]{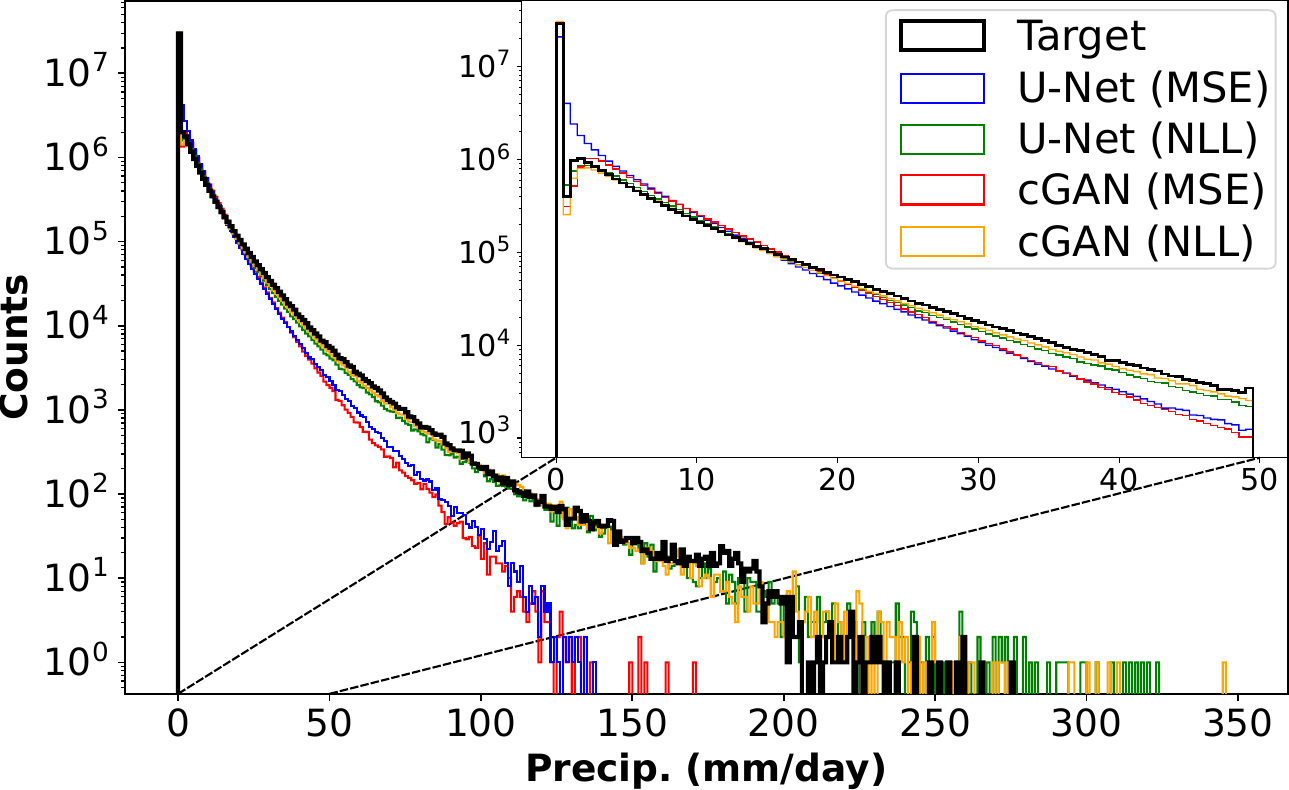}}
        \caption{Histogram of the precipitation distribution for the test period, aggregated across all grid-points in the predictand. The black line represents the target observational dataset, while the different colors correspond to the various DL models being compared. A zoomed-in view for values in the 0-50 interval is provided in the top-right corner of the histogram.}
        \label{fig:histogram}
    \end{center}
\vskip -0.2in
\end{figure}

Figure \ref{fig:histogram} displays the precipitation histogram during the test period across all grid-points in the predictand. The logarithmic scale of the y-axis facilitates the assessment of model performance, particularly for less frequent extreme events. Examining the histogram within the 0-50 mm interval (top-right corner), we observe a notable decline in precipitation values from 0 to approximately 1 mm in the target dataset. The U-Net (MSE) model fails to adapt to this pattern, resulting in an overestimation of precipitation. In contrast, other models accurately replicate this decrease. This discrepancy supports the U-Net (MSE) model's overestimation of the mean in the corresponding violin plot. The adversarial training of the cGAN (MSE) allows it to adjust to this decrease, while NLL-based models effectively handle it due to the underlying distributional assumption. Additionally, the histogram reveals that MSE-based models underestimate the distribution beyond approximately 15 mm, as minimizing the MSE loss function mainly involves fitting the mean. Conversely, NLL-based models accurately reproduce this segment of the distribution, as also evidenced in the SDII violin plot.

Expanding our focus to the histogram spanning the 0-350 mm interval, representing precipitation extremes, a similar pattern emerges: only NLL-based models successfully replicate these extreme values, including those surpassing 200 mm. This underscores the effectiveness of assuming a gamma distribution for modeling the precipitation amount.

\begin{figure}[ht]
\vskip 0.2in
    \begin{center}
        \centerline{\includegraphics[width=0.82\columnwidth]{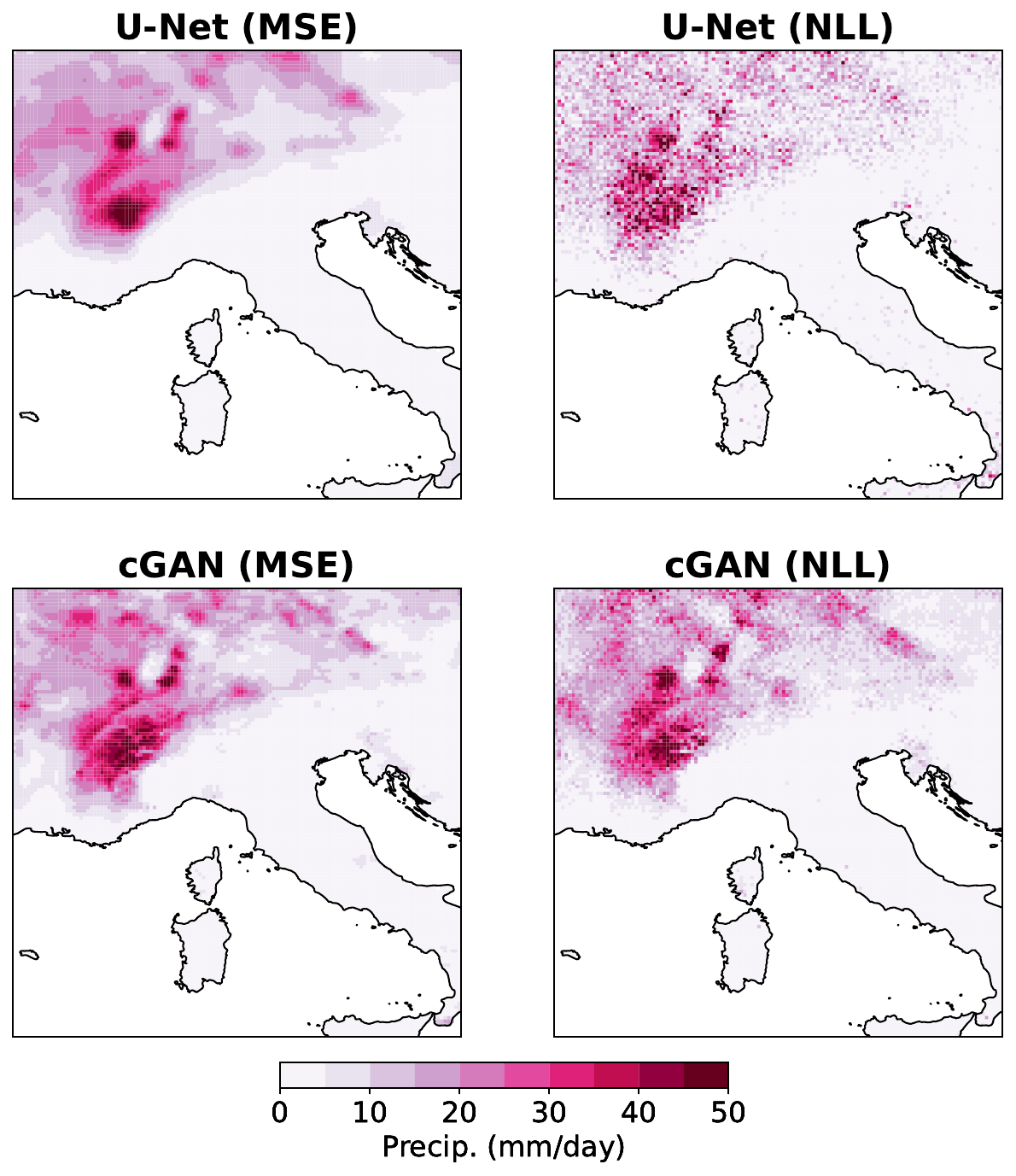}}
        \caption{Comparison of predictions generated by the DL models intercompared for a day in the test period.}
        \label{fig:fine_detail}
    \end{center}
\vskip -0.2in
\end{figure}

Figure \ref{fig:fine_detail} depicts the predictions of each DL model for a specific day of the test period. As anticipated, the U-Net (MSE) exhibits the \textit{blurry effect}, lacking fine details in the generated field. In contrast, the cGAN (MSE) addresses this issue by leveraging the adversarial loss, leading the generator to better capture these details to fool the discriminator, along with the effect of computing the MSE over the ensemble of predictions. As previously discussed, the independent nature of the distributions resulting from minimizing the NLL loss function leads to spatial inconsistencies in the generated field, evident in the U-Net (NLL) model's prediction. However, incorporating the adversarial loss enables the cGAN (NLL) to achieve greater consistency while still adhering to the Bernoulli-gamma distribution. In fact, the prediction of the cGAN (NLL) model falls between that of the cGAN (MSE) and the U-Net (NLL).

\section{Conclusions}

In this work, we have introduced a novel likelihood-based generative approach for precipitation downscaling. This method leverages a combination of likelihood and adversarial losses, enabling the model to properly reproduce the target distribution while generating spatially consistent precipitation fields, addressing a main challenge for standard likelihood-based DL models. Furthermore, likelihood-based loss functions enable generative models to produce explicit probability distributions (e.g., Bernoulli-gamma) for precipitation. This capability is crucial when downscaling future GCM projections, as estimating probabilities of extreme events is vital for risk assessment.

Future work includes evaluating the likelihood-based generative model in the GCM space and compare the resultant projections with those of established PP-SD models. We also plan to employ eXplainable Artificial Intelligence (XAI) techniques to understand the impact of the adversarial loss on the modeled probability distributions and learned patterns.

\section*{Acknowledgements}

We acknowledge support from grant CPP2021-008510 funded by MICIU/AEI/10.13039/501100011033 and by
the “European Union” and the “European Union NextGenerationEU/PRTR”, as well as Project COMPOUND (TED2021-131334A-I00) funded by MCIU/AEI/10.13039/501100011033 and by the European Union NextGenerationEU/PRTR.


\bibliography{references}

\begin{thebibliography}{42}
\providecommand{\natexlab}[1]{#1}
\providecommand{\url}[1]{\texttt{#1}}
\expandafter\ifx\csname urlstyle\endcsname\relax
  \providecommand{\doi}[1]{doi: #1}\else
  \providecommand{\doi}{doi: \begingroup \urlstyle{rm}\Url}\fi

\bibitem[Adewoyin et~al.(2021)Adewoyin, Dueben, Watson, He, and Dutta]{adewoyin_tru_2021}
Adewoyin, R.~A., Dueben, P., Watson, P., He, Y., and Dutta, R.
\newblock Tru-net: a deep learning approach to high resolution prediction of rainfall.
\newblock \emph{Machine Learning}, 110:\penalty0 2035--2062, 2021.

\bibitem[Arjovsky et~al.(2017)Arjovsky, Chintala, and Bottou]{arjovsky_wasserstein_2017}
Arjovsky, M., Chintala, S., and Bottou, L.
\newblock Wasserstein generative adversarial networks.
\newblock In \emph{International conference on machine learning}, pp.\  214--223. PMLR, 2017.

\bibitem[Ba{\~n}o-Medina et~al.(2020)Ba{\~n}o-Medina, Manzanas, and Guti{\'e}rrez]{bano_configuration_2020}
Ba{\~n}o-Medina, J., Manzanas, R., and Guti{\'e}rrez, J.~M.
\newblock Configuration and intercomparison of deep learning neural models for statistical downscaling.
\newblock \emph{Geoscientific Model Development}, 13\penalty0 (4):\penalty0 2109--2124, 2020.

\bibitem[Ba{\~n}o-Medina et~al.(2021)Ba{\~n}o-Medina, Manzanas, and Guti{\'e}rrez]{bano_suitability_2021}
Ba{\~n}o-Medina, J., Manzanas, R., and Guti{\'e}rrez, J.~M.
\newblock On the suitability of deep convolutional neural networks for continental-wide downscaling of climate change projections.
\newblock \emph{Climate Dynamics}, 57\penalty0 (11):\penalty0 2941--2951, 2021.

\bibitem[Ba{\~n}o-Medina et~al.(2022)Ba{\~n}o-Medina, Manzanas, Cimadevilla, Fern{\'a}ndez, Gonz{\'a}lez-Abad, Cofi{\~n}o, and Guti{\'e}rrez]{bano_downscaling_2022}
Ba{\~n}o-Medina, J., Manzanas, R., Cimadevilla, E., Fern{\'a}ndez, J., Gonz{\'a}lez-Abad, J., Cofi{\~n}o, A.~S., and Guti{\'e}rrez, J.~M.
\newblock Downscaling multi-model climate projection ensembles with deep learning (deepesd): contribution to cordex eur-44.
\newblock \emph{Geoscientific Model Development Discussions}, 2022:\penalty0 1--14, 2022.

\bibitem[Cannon(2008)]{cannon_probabilistic_2008}
Cannon, A.~J.
\newblock Probabilistic multisite precipitation downscaling by an expanded bernoulli--gamma density network.
\newblock \emph{Journal of Hydrometeorology}, 9\penalty0 (6):\penalty0 1284--1300, 2008.

\bibitem[Chen et~al.(2021)Chen, Rojas, Samset, Cobb, Diongue~Niang, Edwards, Emori, Faria, Hawkins, Hope, Huybrechts, Meinshausen, Mustafa, Plattner, and Tréguier]{chen_framing_2021-1}
Chen, D., Rojas, M., Samset, B., Cobb, K., Diongue~Niang, A., Edwards, P., Emori, S., Faria, S., Hawkins, E., Hope, P., Huybrechts, P., Meinshausen, M., Mustafa, S., Plattner, G.-K., and Tréguier, A.-M.
\newblock Framing, {Context}, and {Methods}.
\newblock In Masson-Delmotte, V., Zhai, P., Pirani, A., Connors, S., Péan, C., Berger, S., Caud, N., Chen, Y., Goldfarb, L., Gomis, M., Huang, M., Leitzell, K., Lonnoy, E., Matthews, J., Maycock, T., Waterfield, T., Yelekçi, O., Yu, R., and Zhou, B. (eds.), \emph{Climate {Change} 2021: {The} {Physical} {Science} {Basis}. {Contribution} of {Working} {Group} {I} to the {Sixth} {Assessment} {Report} of the {Intergovernmental} {Panel} on {Climate} {Change}}, pp.\  147--286. Cambridge University Press, Cambridge, United Kingdom and New York, NY, USA, 2021.

\bibitem[Cheng et~al.(2020{\natexlab{a}})Cheng, Kuang, Shen, Liu, Tan, and Liu]{cheng_reslap_2020}
Cheng, J., Kuang, Q., Shen, C., Liu, J., Tan, X., and Liu, W.
\newblock Reslap: Generating high-resolution climate prediction through image super-resolution.
\newblock \emph{IEEE Access}, 8:\penalty0 39623--39634, 2020{\natexlab{a}}.

\bibitem[Cheng et~al.(2020{\natexlab{b}})Cheng, Liu, Xu, Shen, and Kuang]{cheng_generating_2020}
Cheng, J., Liu, J., Xu, Z., Shen, C., and Kuang, Q.
\newblock Generating high-resolution climate prediction through generative adversarial network.
\newblock \emph{Procedia Computer Science}, 174:\penalty0 123--127, 2020{\natexlab{b}}.

\bibitem[Cornes et~al.(2018)Cornes, van~der Schrier, van~den Besselaar, and Jones]{cornes_ensemble_2018}
Cornes, R.~C., van~der Schrier, G., van~den Besselaar, E.~J., and Jones, P.~D.
\newblock An ensemble version of the e-obs temperature and precipitation data sets.
\newblock \emph{Journal of Geophysical Research: Atmospheres}, 123\penalty0 (17):\penalty0 9391--9409, 2018.

\bibitem[Doury et~al.(2023)Doury, Somot, Gadat, Ribes, and Corre]{doury_regional_2023}
Doury, A., Somot, S., Gadat, S., Ribes, A., and Corre, L.
\newblock Regional climate model emulator based on deep learning: Concept and first evaluation of a novel hybrid downscaling approach.
\newblock \emph{Climate Dynamics}, 60\penalty0 (5):\penalty0 1751--1779, 2023.

\bibitem[Doury et~al.(2024)Doury, Somot, and Gadat]{doury_suitability_2024}
Doury, A., Somot, S., and Gadat, S.
\newblock On the suitability of a convolutional neural network based rcm-emulator for fine spatio-temporal precipitation.
\newblock \emph{Toulouse School of Economics Repository}, 2024.

\bibitem[Dunn(2004)]{dunn_occurrence_2004}
Dunn, P.~K.
\newblock Occurrence and quantity of precipitation can be modelled simultaneously.
\newblock \emph{International Journal of Climatology: A Journal of the Royal Meteorological Society}, 24\penalty0 (10):\penalty0 1231--1239, 2004.

\bibitem[Gonz{\'a}lez-Abad et~al.(2021)Gonz{\'a}lez-Abad, Ba{\~n}o-Medina, and Cach{\'a}]{gonzalez_use_2021}
Gonz{\'a}lez-Abad, J., Ba{\~n}o-Medina, J., and Cach{\'a}, I.~H.
\newblock On the use of deep generative models for perfect prognosis climate downscaling.
\newblock \emph{arXiv preprint arXiv:2305.00974}, 2021.

\bibitem[Goodfellow et~al.(2014)Goodfellow, Pouget-Abadie, Mirza, Xu, Warde-Farley, Ozair, Courville, and Bengio]{goodfellow_generative_2014}
Goodfellow, I., Pouget-Abadie, J., Mirza, M., Xu, B., Warde-Farley, D., Ozair, S., Courville, A., and Bengio, Y.
\newblock Generative adversarial nets.
\newblock \emph{Advances in neural information processing systems}, 27, 2014.

\bibitem[Goodfellow et~al.(2016)Goodfellow, Bengio, and Courville]{goodfellow_deep_2016}
Goodfellow, I., Bengio, Y., and Courville, A.
\newblock \emph{Deep learning}.
\newblock MIT press, 2016.

\bibitem[Gulrajani et~al.(2017)Gulrajani, Ahmed, Arjovsky, Dumoulin, and Courville]{gulrajani_improved_2017}
Gulrajani, I., Ahmed, F., Arjovsky, M., Dumoulin, V., and Courville, A.~C.
\newblock Improved training of wasserstein gans.
\newblock \emph{Advances in neural information processing systems}, 30, 2017.

\bibitem[Harris et~al.(2022)Harris, McRae, Chantry, Dueben, and Palmer]{harris_generative_2022}
Harris, L., McRae, A.~T., Chantry, M., Dueben, P.~D., and Palmer, T.~N.
\newblock A generative deep learning approach to stochastic downscaling of precipitation forecasts.
\newblock \emph{Journal of Advances in Modeling Earth Systems}, 14\penalty0 (10):\penalty0 e2022MS003120, 2022.

\bibitem[Hersbach et~al.(2020)Hersbach, Bell, Berrisford, Hirahara, Hor{\'a}nyi, Mu{\~n}oz-Sabater, Nicolas, Peubey, Radu, Schepers, et~al.]{hersbach_era5_2020}
Hersbach, H., Bell, B., Berrisford, P., Hirahara, S., Hor{\'a}nyi, A., Mu{\~n}oz-Sabater, J., Nicolas, J., Peubey, C., Radu, R., Schepers, D., et~al.
\newblock The era5 global reanalysis.
\newblock \emph{Quarterly Journal of the Royal Meteorological Society}, 146\penalty0 (730):\penalty0 1999--2049, 2020.

\bibitem[Hosseini~Baghanam et~al.(2024)Hosseini~Baghanam, Nourani, Bejani, and Ke]{hosseini_improving_2024}
Hosseini~Baghanam, A., Nourani, V., Bejani, M., and Ke, C.-Q.
\newblock Improving the statistical downscaling performance of climatic parameters with convolutional neural networks.
\newblock \emph{Journal of Water and Climate Change}, pp.\  jwc2024592, 2024.

\bibitem[Kheir et~al.(2023)Kheir, Elnashar, Mosad, and Govind]{kheir_improved_2023}
Kheir, A.~M., Elnashar, A., Mosad, A., and Govind, A.
\newblock An improved deep learning procedure for statistical downscaling of climate data.
\newblock \emph{Heliyon}, 9\penalty0 (7), 2023.

\bibitem[Kingma \& Ba(2014)Kingma and Ba]{kingma_adam_2014}
Kingma, D.~P. and Ba, J.
\newblock Adam: A method for stochastic optimization.
\newblock \emph{arXiv preprint arXiv:1412.6980}, 2014.

\bibitem[Leinonen et~al.(2020)Leinonen, Nerini, and Berne]{leinonen_stochastic_2020}
Leinonen, J., Nerini, D., and Berne, A.
\newblock Stochastic super-resolution for downscaling time-evolving atmospheric fields with a generative adversarial network.
\newblock \emph{IEEE Transactions on Geoscience and Remote Sensing}, 59\penalty0 (9):\penalty0 7211--7223, 2020.

\bibitem[Maraun \& Widmann(2018)Maraun and Widmann]{maraun_statistical_2018}
Maraun, D. and Widmann, M.
\newblock \emph{Statistical downscaling and bias correction for climate research}.
\newblock Cambridge University Press, 2018.

\bibitem[Mirza \& Osindero(2014)Mirza and Osindero]{mirza_conditional_2014}
Mirza, M. and Osindero, S.
\newblock Conditional generative adversarial nets.
\newblock \emph{arXiv preprint arXiv:1411.1784}, 2014.

\bibitem[Misra et~al.(2018)Misra, Sarkar, and Mitra]{misra_statistical_2018}
Misra, S., Sarkar, S., and Mitra, P.
\newblock Statistical downscaling of precipitation using long short-term memory recurrent neural networks.
\newblock \emph{Theoretical and applied climatology}, 134:\penalty0 1179--1196, 2018.

\bibitem[O'Neill et~al.(2016)O'Neill, Tebaldi, Van~Vuuren, Eyring, Friedlingstein, Hurtt, Knutti, Kriegler, Lamarque, Lowe, et~al.]{o_scenario_2016}
O'Neill, B.~C., Tebaldi, C., Van~Vuuren, D.~P., Eyring, V., Friedlingstein, P., Hurtt, G., Knutti, R., Kriegler, E., Lamarque, J.-F., Lowe, J., et~al.
\newblock The scenario model intercomparison project (scenariomip) for cmip6.
\newblock \emph{Geoscientific Model Development}, 9\penalty0 (9):\penalty0 3461--3482, 2016.

\bibitem[Pan et~al.(2019)Pan, Hsu, AghaKouchak, and Sorooshian]{pan_improving_2019}
Pan, B., Hsu, K., AghaKouchak, A., and Sorooshian, S.
\newblock Improving precipitation estimation using convolutional neural network.
\newblock \emph{Water Resources Research}, 55\penalty0 (3):\penalty0 2301--2321, 2019.

\bibitem[Passarella et~al.(2022)Passarella, Mahajan, Pal, and Norman]{passarella_reconstructing_2022}
Passarella, L.~S., Mahajan, S., Pal, A., and Norman, M.~R.
\newblock Reconstructing high resolution esm data through a novel fast super resolution convolutional neural network (fsrcnn).
\newblock \emph{Geophysical Research Letters}, 49\penalty0 (4):\penalty0 e2021GL097571, 2022.

\bibitem[Price \& Rasp(2022)Price and Rasp]{price_increasing_2022}
Price, I. and Rasp, S.
\newblock Increasing the accuracy and resolution of precipitation forecasts using deep generative models.
\newblock In \emph{International conference on artificial intelligence and statistics}, pp.\  10555--10571. PMLR, 2022.

\bibitem[Quesada-Chac{\'o}n et~al.(2022)Quesada-Chac{\'o}n, Barfus, and Bernhofer]{quesada_repeatable_2022}
Quesada-Chac{\'o}n, D., Barfus, K., and Bernhofer, C.
\newblock Repeatable high-resolution statistical downscaling through deep learning.
\newblock \emph{Geoscientific Model Development}, 15\penalty0 (19):\penalty0 7353--7370, 2022.

\bibitem[Rampal et~al.(2022)Rampal, Gibson, Sood, Stuart, Fauchereau, Brandolino, Noll, and Meyers]{rampal_high_2022}
Rampal, N., Gibson, P.~B., Sood, A., Stuart, S., Fauchereau, N.~C., Brandolino, C., Noll, B., and Meyers, T.
\newblock High-resolution downscaling with interpretable deep learning: Rainfall extremes over new zealand.
\newblock \emph{Weather and Climate Extremes}, 38:\penalty0 100525, 2022.

\bibitem[Rampal et~al.(2024{\natexlab{a}})Rampal, Gibson, Sherwood, Abramowitz, and Hobeichi]{rampal_robust_2024}
Rampal, N., Gibson, P.~B., Sherwood, S., Abramowitz, G., and Hobeichi, S.
\newblock A robust generative adversarial network approach for climate downscaling and weather generation.
\newblock \emph{Authorea Preprints}, 2024{\natexlab{a}}.

\bibitem[Rampal et~al.(2024{\natexlab{b}})Rampal, Hobeichi, Gibson, Ba{\~n}o-Medina, Abramowitz, Beucler, Gonz{\'a}lez-Abad, Chapman, Harder, and Guti{\'e}rrez]{rampal_enhancing_2024}
Rampal, N., Hobeichi, S., Gibson, P.~B., Ba{\~n}o-Medina, J., Abramowitz, G., Beucler, T., Gonz{\'a}lez-Abad, J., Chapman, W., Harder, P., and Guti{\'e}rrez, J.~M.
\newblock Enhancing regional climate downscaling through advances in machine learning.
\newblock \emph{Artificial Intelligence for the Earth Systems}, 3\penalty0 (2):\penalty0 230066, 2024{\natexlab{b}}.

\bibitem[Ravuri et~al.(2021)Ravuri, Lenc, Willson, Kangin, Lam, Mirowski, Fitzsimons, Athanassiadou, Kashem, Madge, et~al.]{ravuri_skilful_2021}
Ravuri, S., Lenc, K., Willson, M., Kangin, D., Lam, R., Mirowski, P., Fitzsimons, M., Athanassiadou, M., Kashem, S., Madge, S., et~al.
\newblock Skilful precipitation nowcasting using deep generative models of radar.
\newblock \emph{Nature}, 597\penalty0 (7878):\penalty0 672--677, 2021.

\bibitem[Ronneberger et~al.(2015)Ronneberger, Fischer, and Brox]{ronneberger_unet_2015}
Ronneberger, O., Fischer, P., and Brox, T.
\newblock U-net: Convolutional networks for biomedical image segmentation.
\newblock In \emph{Medical image computing and computer-assisted intervention--MICCAI 2015: 18th international conference, Munich, Germany, October 5-9, 2015, proceedings, part III 18}, pp.\  234--241. Springer, 2015.

\bibitem[Sharma \& Mitra(2022)Sharma and Mitra]{sharma_resdeepd_2022}
Sharma, S. C.~M. and Mitra, A.
\newblock Resdeepd: A residual super-resolution network for deep downscaling of daily precipitation over india.
\newblock \emph{Environmental Data Science}, 1:\penalty0 e19, 2022.

\bibitem[Soares et~al.(2023)Soares, Johannsen, Lima, Lemos, Bento, and Bushenkova]{soares_high_2023}
Soares, P.~M., Johannsen, F., Lima, D.~C., Lemos, G., Bento, V., and Bushenkova, A.
\newblock High resolution downscaling of cmip6 earth system and global climate models using deep learning for iberia.
\newblock \emph{Geoscientific Model Development Discussions}, 2023:\penalty0 1--46, 2023.

\bibitem[Sun \& Lan(2021)Sun and Lan]{sun_statistical_2021}
Sun, L. and Lan, Y.
\newblock Statistical downscaling of daily temperature and precipitation over china using deep learning neural models: Localization and comparison with other methods.
\newblock \emph{International Journal of Climatology}, 41\penalty0 (2):\penalty0 1128--1147, 2021.

\bibitem[Vandal et~al.(2017)Vandal, Kodra, Ganguly, Michaelis, Nemani, and Ganguly]{vandal_deepsd_2017}
Vandal, T., Kodra, E., Ganguly, S., Michaelis, A., Nemani, R., and Ganguly, A.~R.
\newblock Deepsd: Generating high resolution climate change projections through single image super-resolution.
\newblock In \emph{Proceedings of the 23rd acm sigkdd international conference on knowledge discovery and data mining}, pp.\  1663--1672, 2017.

\bibitem[Wang et~al.(2020)Wang, Chen, and Hoi]{wang_deep_2020}
Wang, Z., Chen, J., and Hoi, S.~C.
\newblock Deep learning for image super-resolution: A survey.
\newblock \emph{IEEE transactions on pattern analysis and machine intelligence}, 43\penalty0 (10):\penalty0 3365--3387, 2020.

\bibitem[Williams(1997)]{williams_modelling_1997}
Williams, P.
\newblock Modelling {Seasonality} and {Trends} in {Daily} {Rainfall} {Data}.
\newblock In \emph{Advances in {Neural} {Information} {Processing} {Systems}}, volume~10. MIT Press, 1997.

\end{thebibliography}
\bibliographystyle{icml2024}




\end{document}